# Communication quality in extreme environments affects performance of astronauts and their support teams through increases in workload: Insights from the AMADEE-20 analog Mars mission

Vera Hagemann [a*], Lara Watermann [a], Florian Klonek [b] and Christiane Heinicke [c]

[a] Chair of Business Psychology & Human Resource Management, Faculty of Business Studies and Economics, University of Bremen, Enrique-Schmidt-Strasse 1, 28359 Bremen, Germany

[b] Centre for Transformative Work Design, Faculty of Business and Law, Curtin University, 78 Murray Street, Perth WA 6000, Australia

[c] Center of Applied Space Technology and Microgravity – ZARM, University of Bremen, Am Fallturm 2, 28359, Bremen, Germany

* Corresponding author.

*E-mail addresses:* vhagemann@uni-bremen.de (V. Hagemann), la_wa@uni-bremen.de (L. Watermann), florian.klonek@curtin.edu.au (F. Klonek), christiane.heinicke@zarm.uni-bremen.de (C. Heinicke).

**Abstract**

Astronaut crews and ground control support teams are highly interdependent teams that need to communicate effectively to achieve a safe mission - despite being separated by large distances. Problems in communication have been shown to lead to increasing effort and frustration in teams working under time-delayed communication. Team communication quality with its facets clarity of objectives and information flow, is a key coordination process to achieve high team performance and task satisfaction. Especially in interdependent teams working in extreme environments with time-delayed communications, the team's success is threatened if communication is ineffective. In this study, we hypothesized that communication quality (clarity of objectives and information flow) affects two key team outcomes, performance and task satisfaction, and that these effects can be explained by increases in workload (effort and frustration). Hypotheses were tested during the AMADEE-20 analog Mars mission hosted by the Austrian Space Forum during which a crew of six analog astronauts (AAs) lived in isolation in the Negev desert in Israel over three weeks. The astronauts were supported by an On-Site-Support (OSS) team and a remote Mission-Support-Centre (MSC) team. The MSC was the only contact line for both AA and OSS, and the communication between them had a one-way time delay of 10 minutes. There was no direct communication between AAs and OSS, despite their spacial proximity. Our study consisted of three runs that involved inter-team coordination between members across the three different multiteam systems: Two participants each from the AAs, MSC and OSS worked together as a multi-team six-person system distributed over three separate locations and had to exchange information to solve an interdependent task within a specified time. We measured communication quality, workload (i.e., effort and frustration), task satisfaction, and team performance. Results show that clarity of objectives and information flow positively impacted multiteam system performance. Furthermore, clarity of objectives reduced experienced effort and this in turn enhances team performance. High levels of information flow, on the other hand, reduced experienced frustration, which in turn enhanced task satisfaction. Our findings show that these facets of communication quality are essential for multiteam systems that work interpedently but are separated from each other by a distance. We stress that specific (team) communication training for astronauts and support personnel will be key to effective teamwork during future Mars missions, and thus to overall mission success.

# 1. Introduction

Travelling to Mars will not only bring new scientific insights, it first of all presents astronauts and their support teams with many unknown risks and challenges. A major challenge for future exploration missions to destinations beyond cis-lunar space will be the communication with Earth: The delay time in communication between Earth and Mars can increase up to 22 minutes one-way [1,2].

Nevertheless, mission-relevant information sharing between astronauts and ground support is essential for task execution, up-to-date shared situation awareness and shared mental models as the mission evolves [1]. Situation awareness and shared mental models are important as they enable team members to develop knowledge structures, which in return facilitate the development of precise expectations to plan their actions and modify their behaviour [3]. Furthermore, knowledge and expertise are distributed across astronauts and their support teams making communication and teamwork indispensable as mission goals are only attained by a joint effort [1]. Therefore, astronauts as well as their support teams must collaborate as a multiteam system, which is defined as two or more teams collaboratively working together towards a shared goal, to execute a variety of tasks [4]. They have to troubleshoot vehicle maintenance issues, conduct on-board experimental research, and make on-the-fly resource decisions additionally to their pre-scheduled daily duties. These responsibilities will require those teams to collaborate and think in inventive and novel ways, effectively combine knowledge resources, and make morally difficult choices to face the unknowns that the teams will undoubtedly experience in the future of long-duration space missions [5]. In short, the inter-team communication of astronauts and their support teams is essential for mission success.

Therefore, the aim of this study was first to identify important facets of communication quality and to highlight their influence on multiteam system performance and task satisfaction in the presence of delayed communication. Second, we aim to show that the impact of these two facets (i.e., information flow and clarity of objectives) on performance and task satisfaction can be explained by the frustration and effort experienced during the inter-team collaboration. This paper concludes with recommendations on how to prevent the negative effects of time-delayed communication between interdependent team members in order to make long-duration space missions more successful.

## 1.1 Teamwork in long-duration spaceflight

The success of future long-duration space missions will depend on effective teamwork [6], especially on the effectiveness of multiteam systems [4]. This is supported by experienced astronauts and other spaceflight subject matter experts as they assessed teamwork and competencies promoting team functioning as essential factors for the success of future long-duration space missions. They highlight that teamwork is vital to mission success, regardless of whether astronauts spend three months on the International Space Station (ISS) or three years on a Mars mission [6].

A mission to Mars requires astronauts and their support teams to work under extreme conditions that present both high physical and psychological risks and challenges. Therefore, going to Mars demands teams capable of enduring and sustaining unprecedented levels of team performance [7]. In particular, team members are required to cooperate, communicate, and coordinate for longer time periods while working in challenging situations and under extreme and highly uncertain conditions and surroundings, e.g., isolation and unknown territory [7,8,9]. Faults and mistakes in the work of those teams can lead to severe, even life-threatening consequences if they are not detected and corrected as soon as possible [8,9]. To avoid or quickly recognize mistakes, the teams need to manage the complexity of the given situations and stay focused as well as flexible at the same time. Taking it all together, both astronauts and their support teams perceive a high mental workload due to extreme information processing demands and belong to the group of extreme teams [8] or high responsibility teams (HRTs) [9]. HRTs are not only characterized by complex working conditions, but also by complex relationships between team members, so called crew coordination complexity [10]. Thus, there are enormous coordination demands in HRTs because a high number of interactive, continuous processes have to be coordinated to successfully perform as HRTs [11,12]. Furthermore, situations change quickly, critical information may be absent, incorrect, or misleading by the time the team receives it, and teams do not always get immediate feedback on how effective their efforts are. Nevertheless, the teams are obligated to predict future states while planning their actions and are required to achieve many, sometimes contradictory goals at the same time [9,13]. This again highlights the importance of effective inter-team communication between astronauts and their support teams to keep the mental effort and the experienced frustration as low as possible. In the following the term multiteam system includes astronauts and their support teams.

## 1.2 Communication between astronauts and their support teams in long-duration spaceflight

Successful team performance requires teams to adapt their communication, reprioritize their objectives as well as planned actions and to reallocate their workload dynamically [14,15]. This becomes more challenging when team members within a multiteam system have to collaborate over massive distances, which means that communication between team members has a significant time delay (i.e., several minutes). To support successful inter-team communication, chatting software is used to coordinate activities [16]. However, establishing a mutual understanding is more effortful, as time-delayed communication comes with considerable cognitive load. Misunderstandings are more likely in delayed communication as there is no immediate feedback and resolving misunderstandings requires a lot of time and workload in form of mental effort [17,18,19]. In addition, the occurrence of misunderstandings in communication is more likely when personnel rotate in ground support, as they do in reality [20]. A study by Kintz et al. [21] showed that the well-being of team members significantly decreased when having a 50-second one-way communication delay compared to when having no communication delay. Next to this, the team's task efficiency, coordination as well as individual and organizational outcomes, i.e., job involvement and satisfaction, were affected, and team members experienced more frustration when misunderstandings occurred [21]. To avoid losing track of conversations and prevent misunderstandings, especially in episodes of frequent conversations, team members typically include more detail in their messages to guarantee that their messages are not only received but also understood [16, 20]. These findings are in line with the Source-Message-Channel-Receiver model of communication as the channel is limited since interpersonal characteristics of communication and immediate feedback are unfeasible in time-delayed communication [22]. Thus, information needs to circulate continuously between transmitter and receiver to be able to understand each other correctly. This implies that the *information flow* is an important factor when considering the communication of time-delayed teams. Information flow means that relevant information is continuously exchanged across team members and that their information is not disturbed so that information is linked together continually [23]. Disturbed information flow leads to higher workload in form of high *frustration*, which in return has a negative impact on organizational outcomes [24]. To date no study has yet investigated how information flow in delayed communication in the context of spaceflight leads to organizational and individual outcomes, e.g., performance and task satisfaction, and the role of frustration within this relationship.

In addition, it was demonstrated that time-delayed communication is successful when the teams adapt to the restraints of their communication and establish a shared understanding [17]. This was also supported by Olson and Olson [25] who showed that team members that work geographically separated from each other need a high common understanding for successful team performance, underlying that clarifying objectives is essential to achieve a shared understanding. These findings highlight the importance of *clarifying the objectives* in time delayed communication. This is in line with the theory of goal setting and task performance by Locke and Latham [26], which indicates that individuals who have specific and clear goals perform better compared to individuals that set vague and unclear goals. The theory goes further by incorporating mediators, i.e., workload in terms of *effort*, that influence the relationship between goal setting and task performance [26]. Based on this, it can be assumed that clarity of objectives is an important facet of communication quality, which has an influence on team performance and that this influence is mediated by workload in terms of mental effort.

**1.3 The Present Study**

During the AMADEE-20 analog Mars mission in October 2021, we had the unique opportunity to examine the teamwork of interdependently and time-delayed working teams in the context of human spaceflight. It is the aim of this study to investigate facets which are important in successful time-delayed communication, namely clarity of objectives and information flow, in relation to workload (effort and frustration), team performance and task satisfaction in the teamwork of the astronauts and their support teams. Figure 1 provides an overview of the variables of interest in this study classified into independent, mediating, and dependent variables.

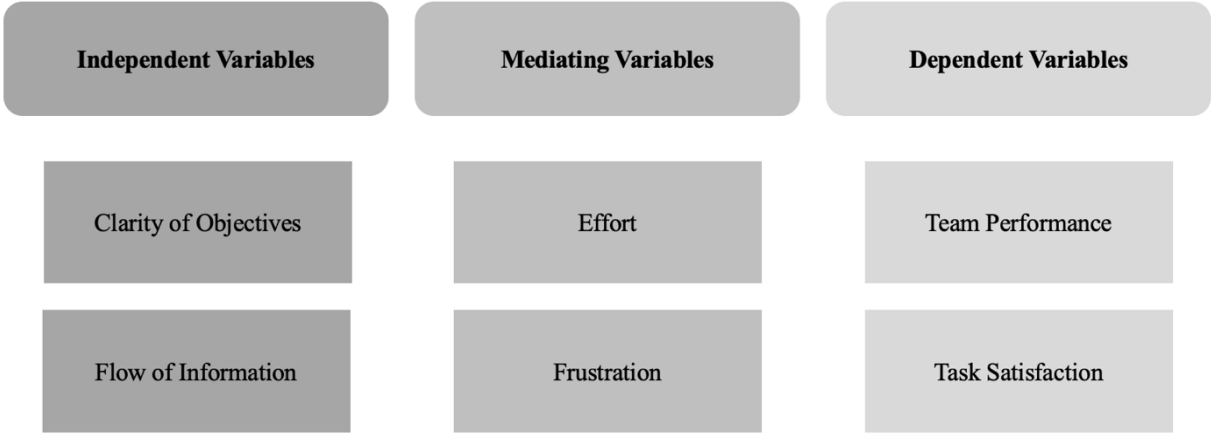

**Figure 1.**
*Overview of independent, mediating and dependent variables studied (own illustration).*

Gaining knowledge in this area is important to prevent communication problems in future exploration missions and to develop training opportunities and technical tools that can support these teams in their communication. Therefore, the following hypotheses are formulated. Firstly, time-delayed communication results in more effort and establishing a shared understanding between teams reduces this effort and leads to successful outcomes [17,25]. Thus, it is expected that the clarity of objectives in time-delayed communication reduces the experienced effort and this in turn leads to higher team performance and to more task satisfaction:

*Hypothesis 1a*: Clarity of objectives is positively associated with team performance.

*Hypothesis 1b*: Effort is negatively associated with team performance.

*Hypothesis 1c*: Effort mediates the relationship between clarity of objectives and team performance.

*Hypothesis 2a*: Clarity of objectives is positively associated with task satisfaction.

*Hypothesis 2b*: Effort is negatively associated with task satisfaction.

*Hypothesis 2c*: Effort mediates the relationship between clarity of objectives and task satisfaction.

Secondly, poor communication and misunderstandings in time-delayed communication are associated with experiencing frustration and less task efficiency and individual outcomes [16,20,21]. Therefore, it is also expected that good information flow in time-delayed communication reduces the experienced frustration and this in turn leads to higher team performance and more task satisfaction:

*Hypothesis 3a*: Information flow is positively associated with team performance.

*Hypothesis 3b*: Frustration is negatively associated with team performance.

*Hypothesis 3c*: Frustration mediates the relationship between information flow and team performance.

*Hypothesis 4a*: Information flow is positively associated with task satisfaction.

*Hypothesis 4b*: Frustration is negatively associated with task satisfaction.

*Hypothesis 4c*: Frustration mediates the relationship between information flow and task satisfaction.

## 2. Material and Methods

**2.1 Design**

A four-week field study design was used to examine communication, workload, team performance and task satisfaction in interdependently and time-delayed working teams in extreme environments. The study was part of the analog Mars mission AMADEE-20, which was hosted by the Austrian Space Forum, the Israeli Space Agency and the organization D-MARS and was conducted from the 4th of October 2021 to the 31st of October 2021 in the Negev Desert, Israel and in Innsbruck, Austria [27].

**2.2 Participants**

The team of the AAs consisted of five men and one woman from Portugal, Spain, Germany, Israel, Austria, and the Netherlands. Their age ranged from 33 to 43. They were chosen from an extensive selection process and ran through a several months-long training, which was specially focused on the Mars spacesuit simulator called Aouda developed by the Austrian Space Forum. During the AMADEE-20 mission the AAs lived isolated in a habitat in the Negev Desert in Israel, which was provided by the organization D-MARS. The MSC in Innsbruck and the OSS team in Israel had fluctuating team members as team members were joining and leaving the teams before and throughout the mission. OSS and MSC consisted each of team members of different age, different nationalities as well as different professional backgrounds. Their age ranged from 18 to 50 and there was an even gender distribution in both teams. To ensure privacy, no further demographics of the participants were collected.

**2.3 The multiteam system**

The multiteam system of AMADEE-20 consisted of (1) a crew of six analog astronauts (AAs) that lived for four weeks in isolation in a habitat in the Negev Desert in Israel, (2) the Mission-Support-Centre (MSC) in Innsbruck, in Austria, which simulated the ground segment of the mission, and (3) the On-Site-Support (OSS) team, which was responsible for the infrastructure in Israel, so that the mission could proceed smoothly. The OSS team is not part of the regular set-up of spaceflight missions, and it consisted of a combination of experienced former analog astronauts and field crew members as well as D-MARS representatives [27]. OSS and AAs were not allowed to communicate with each other. Both teams were coordinated through MSC. During the whole mission there was a 10-minutes time delay between MSC and the other two teams which simulated the average signal traveling time between Earth and Mars.

The time delay also applies to the communication between OSS and MSC as OSS was synchronized with the flight crew [27]. In our study, OSS simulates the remote science support team, as in real missions there is also contact between this team and the ground support. OSS is suitable for this because the members of OSS mostly knew each other, similar to real remote science support teams. Therefore, OSS was included in this study even though they do not belong to the usual spaceflight set-up. Figure 2 depicts the three teams with their respective team members and areas of responsibility as well as the relationships among them.

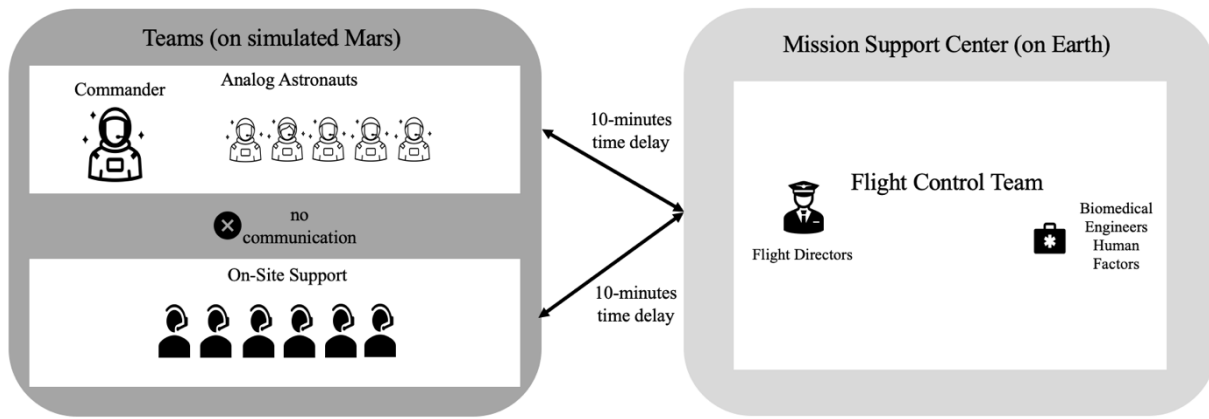

**Figure 2.**
*Overview of the multiteam system of the AMADEE-20 analog Mars mission (own illustration based on Austrian Space Forum [27]).*

## 2.4 Procedure

From the entire sample, we assembled a set of three different multiteam systems. That is, we sampled two participants from each of the sub-teams (i.e., AA, MSC, OSS) and allocated them into a multiteam system – consisting of six participants. None of the participants were allowed to participate twice, that is, each participant was only measured once during the mission. There was one run of this study with one multiteam system per week, so that there were three runs and respectively three different multiteam systems in total. The execution of the task for the multiteam system took approximately 100 minutes including 60 minutes of waiting times due to 10-minutes time delay between Earth and simulated Mars. The astronauts and the support teams had to solve riddles and exchange the solution of the riddles as these were clues for the upcoming riddles and important for the solution in the end. The procedure was the same for all three multiteam systems and is shown in Figure 3 as an example for one study run with one multiteam system.

Simultaneously, two team members from the AAs and two team members from OSS started with reading the instructions for the given task (see top left in Figure 3). Right after, they set a visible timer for five minutes and worked on solving their first riddle. As OSS and the AAs were not able to contact each other, they solved the riddles within their own two-person team and sent the solutions to MSC (see right side in Figure 3). In the meantime, the two team members from MSC started reading the instructions for the given task. When the two team members from MSC received the solutions from both the AAs and OSS, they set a visible timer of five minutes and started solving their first riddle. After the five minutes were over, the team members of MSC sent their solutions to the AAs and to OSS (see middle left in Figure 3). The same procedure was followed two more times. So that each the AAs, OSS and MSC solved three riddles and exchanged the solutions respectively (see Figure 3).

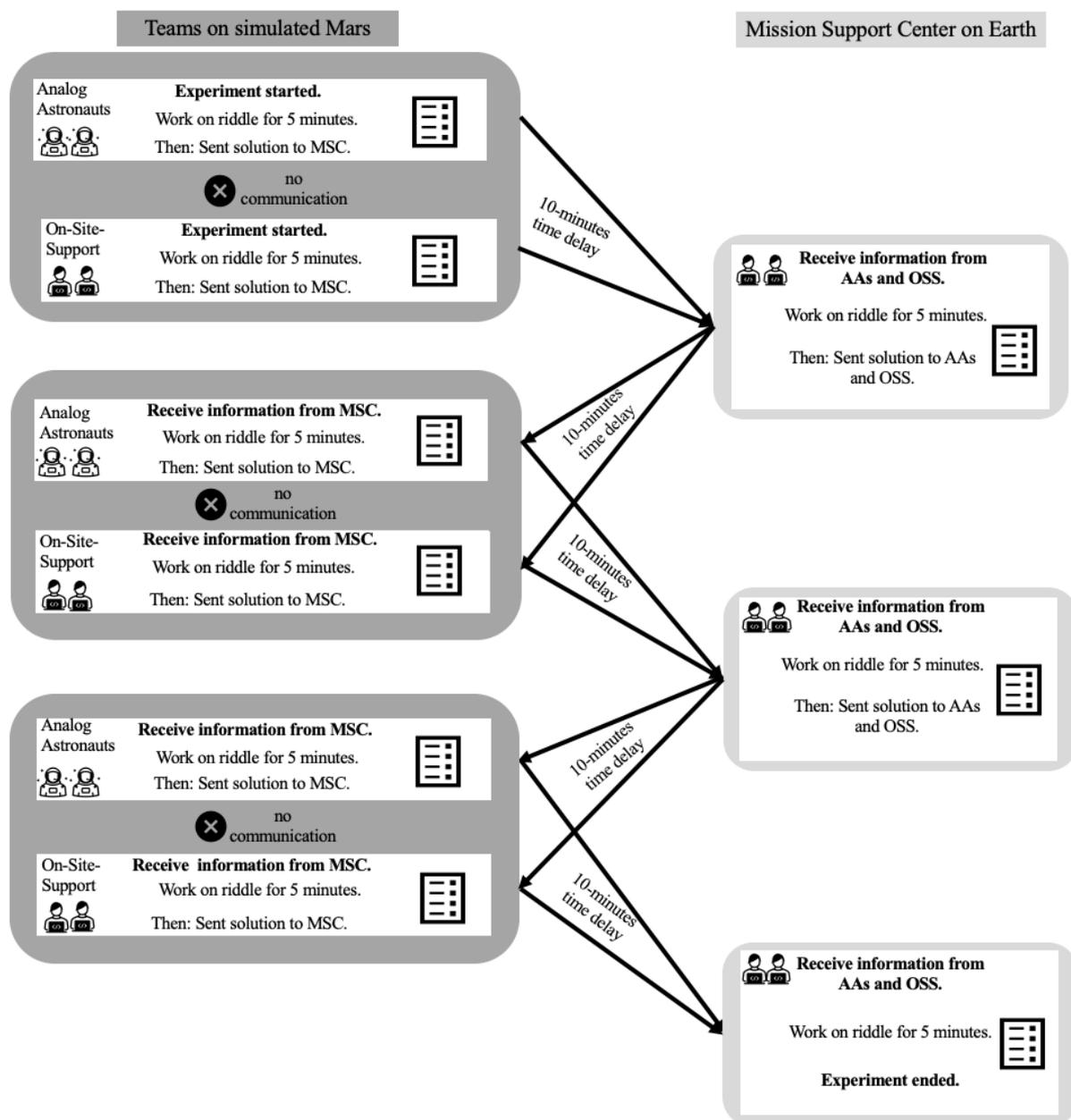

**Figure 3.**
*Study Procedure. Example for a run with a multiteam system with six members (own illustration).*

## 2.5 Team Task

Members of multiteam systems are highly dependent within their multiteam system, so that it is on the one hand important that they clearly communicate all relevant information. On the other hand, uncertainty, and autonomy forces team members in the context of spaceflight to use their knowledge to solve situations that are uncertain [5]. In an emergency, a considerable time delay requires the crew of astronauts to choose a solution before there is time to communicate with support teams. Therefore, they must be able to successfully communicate their knowledge and skills with one another to reach a quick consensus [5]. Therefore, riddles were chosen as a task for the multiteam system. Riddles require team members to work together interdependently by exchanging important clues and additionally demand team members in the event of missing or incorrect information to make decisions based on their own without consulting the other teams in the multiteam system.

In all three runs, it was the multiteam system's overall task to solve a murder case (different case each time; see Appendix A, Table A.1). Specifically, the two members from the AAs had to solve three riddles, the two members from OSS had to solve three riddles and the two members from MSC had to solve three riddles. The riddles contained information on the person that was murdered, the murder weapon that was used and the murderer. The team members had to share their information with each other to get clues for their own riddles in order solve the murder case correctly (see section 2.4. Procedure). Therefore, one multiteam system had to solve nine riddles in total (with each team within the multiteam system solving three riddles each), which in sum solved the given murder case.

## 2.6 Measures

At the end of the study, all participants had to individually fill in a questionnaire containing measures on the communication quality, workload and task satisfaction.

### 2.6.1 Communication Quality

The communication quality measure was set up by using items of Hirst's and Mann's [28] communication measure including the scales *Clarity of Objectives* (e.g., "All of us in the team had a clear understanding of the task's objectives.") and *Information flow* (e.g., "Information regarding the task circulated through the team.") as well as by using Wauben's et al. [29] Three Categories of communication (see Appendix A, Tables A.2 and A.3) [28,29]. In total, the communication measure had six items with a five-point Likert scale ranging from 1 (never) to 5 (always). The higher the score, the better the communication quality was perceived.

In the current study the communication quality measure showed a very good reliability with Cronbach's alpha of .95, with the subscales *Clarity of Objectives* (α = .82; 2 items) and *Information Flow* (α = .95; 4 items).

*2.6.2 Workload*

To assess the workload in terms of effort and frustration, the NASA-Task Load Index was used [30]. Effort was measured using the one-item scale "How hard did you have to work (mentally and physically) to accomplish your level of performance?" and frustration was assessed using the one-item scale "How insecure, discouraged, irritated, stressed and annoyed versus secure, gratified, content, relaxed and complacent did you feel during the task?". Both scales had a 20-point Likert scale ranging from 1 (very low) to 20 (very high).

*2.6.3 Task Satisfaction*

To measure task satisfaction four items of the Job Satisfaction Survey by Spector [31] were taken and adjusted according to the situation of the analog Mars mission simulation (see Appendix A, Table A.4). To answer these items, a six-point Likert scale ranging from 1 (totally disagree) to 6 (strongly agree) was provided (e.g., "The work during the task was enjoyable."). The higher the score, the higher the task satisfaction of the participants. In the current study the task satisfaction measure showed to have a good reliability with Cronbach's alpha of .87.

*2.6.4 Team Performance*

To assess team performance a score depending on how many riddles the multiteam system had solved correctly was given. There were nine riddles per multiteam system in total including three riddles solved by the two AAs, three riddles solved by the two members of OSS and three riddles solved by the two members of MSC. As we considered the three teams as one multiteam system, there was one multiteam score depending on the riddles that were solved correctly. Since there were nine riddles per multiteam system in total, the highest score for the team performance was nine indicating successful team performance.

**2.7 Statistical Analysis**

All analyses were performed using SPSS, Statistical Package for the Social Sciences, version 27. There were no missing values in the data. Due to the Likert scaling, a metric scale level was assumed for the items. The scale values were formed by calculating the arithmetic mean of the items and standard deviations of the variables were considered. As the variables had different ranges of Likert scales, the z-scores were used for all analyses to test the mediation effect of workload (effort, frustration) on the relationship between both facets of communication quality (clarity of objectives and information flow) and team performance as well as task satisfaction. Four mediation analyses were conducted. In the first and the second mediation analysis clarity of objectives was the independent variable and effort was the mediator. Team performance was the dependent variable in the first mediation analysis (Hypothesis 1) while task satisfaction was the dependent variable in the second mediation analysis (Hypothesis 2). In the third and the fourth mediation analysis information flow was the independent variable and frustration was the mediator. Again, team performance was the dependent variable in the third mediation analysis (Hypothesis 3) while task satisfaction was the dependent variable in the fourth mediation analysis (Hypothesis 4). The macro PROCESSv4.0 for SPSS [32] was used for the mediation analyses. The significance level for all analyses was α = .05 and bootstrapping (based on 5000 iterations) was used in every analysis as the sample size was 18 [33].

## 3. Results

### 3.1 Descriptive Results

The means, standard deviations and correlation coefficients of all study variables can be found in Table 1. As the variables had different ranges of Likert scales, the z-scores were used for all analyses. The most important conditions to be tested were linearity and normality [33]. An analysis of the bivariate scatter plots showed that the variables were approximately linearly related (see Appendix B Figure 1). As the sample size is too small (N = 18) to assume normality, Spearman correlation was used [33].

**Table 1**
*Means, standard deviations and Spearman correlation coefficients of the variables studied.*

| Variable | M | SD | 1 | 2 | 3 | 4 | 5 | 6 |
|---|---|---|---|---|---|---|---|---|
| 1. Clarity of objectives | 4.33 | .71 | - | | | | | |
| 2. Information flow | 4.10 | .90 | .85** | - | | | | |
| 3. Effort | 9.33 | 3.07 | -.62** | -.62** | - | | | |
| 4 Frustration | 8.22 | 5.22 | -.53* | -.68** | .40* | - | | |
| 5. Team performance | 6.67 | 1.75 | .74** | .79** | -.69** | -.51* | - | |
| 6. Task satisfaction | 4.40 | 1.12 | .45* | .58* | -.40* | -.57** | .70** | - |

*Note.* N = 18. * $p < .05$, ** $p < .01$.

Table 1 shows that both facets of communication quality, clarity of objectives and information flow, were significantly and positively correlated. Effort was significantly and negatively correlated with clarity of objectives, information flow, team performance and task satisfaction and positively correlated with frustration. Frustration was significantly and negatively correlated with clarity of objectives and information flow. Additionally, there was a significant negative correlation between frustration and team performance as well as between frustration and task satisfaction. Both clarity of objectives and information flow were significantly and positively correlated to team performance and task satisfaction. Finally, team performance and task satisfaction have been shown to be significantly and positively correlated.

**3.2 Hypothesis Tests**

The test of the mediation hypotheses was based on Baron and Kenny [34], who report a significant correlation between the independent variable and the dependent variable as a prerequisite for mediation, and on Preacher and Hayes [35], who test the significance of an indirect effect using bootstrapping confidence intervals. Hayes' PROCESS macro was used for the calculation. This offered the advantage that it already centered the independent variables, automatically calculated the interaction term and produced a simple slope analysis [33]. In addition, robust standard errors were used in the PROCESS macro and confidence intervals were calculated by bootstrapping. These are considered robust to violations of the mediation analysis prerequisite of normal distribution and homoscedasticity [33]. Therefore, these prerequisites were not tested separately. Only the prerequisite of linearity was checked beforehand using matrix scatter plots. The relationships of all variables involved in the mediation analysis were approximately linear, as assessed by visual inspection of scatterplots (see Appendix B, Figure B.1).

The first two hypotheses proposed that clarity of objectives is positively associated with team performance and effort negatively. The correlation analyses in Table 1 show that there is a significant positive correlation between clarity of objectives and team performance ($\rho = .74$, $p < .01$), supporting H1a. The correlation between effort and team performance is significantly negative ($\rho = -.69$, $p < .01$), supporting H1b.

In H1c it was assumed that clarity of objectives reduces experienced effort and this in turn enhances team performance. The results of the mediation analysis showed that there was a total effect of clarity of objectives on team performance ($B = .62$, $p < .01$) in the assumed direction. After including the mediator into the model, clarity of objectives predicted effort significantly and negatively ($B = -.54$, $p < .01$), and effort in return predicted team performance significantly negatively ($B = -.46$, $p < .05$, indirect effect $ab = .25$, 95%-CI[.05, .48]; see Appendix B, Tables 1 to 5 for a detailed overview of the results). The relationship between clarity of objectives and team performance was not significant after the mediator was entered.

Therefore, it was shown that the relationship between clarity of objectives and team performance was mediated by effort, supporting H1c (see Figure 4).

H2a and H2b proposed that clarity of objectives is positively associated with task satisfaction and effort negatively. The correlation analyses in Table 1 show that there is a significant positive correlation between clarity of objectives and task satisfaction ($\rho = .45$, $p < .05$), supporting H2a. Furthermore, the correlation between effort and task satisfaction is negative and significant ($\rho = -.40$, $p > .05$), supporting H2b.

It was expected in H2c that clarity of objectives reduces experienced effort and this in turn enhances task satisfaction. The results of the mediation analysis showed that there was a total effect of clarity of objectives on task satisfaction ($B = .37$, $p < .05$) in the assumed direction. After including the mediator into the model, clarity of objectives significantly predicted effort negatively ($B = -.54$, $p < .01$), but effort in return did not predict task satisfaction significantly ($B = -.12$, $p > .05$, indirect effect $ab = .07$, 95%-CI[-.23, .43]; see Appendix B, Tables B.2 to B.3 and Tables B.6 to B.8 for a detailed overview of the results). Thus, H2c was not supported (see Figure 4).

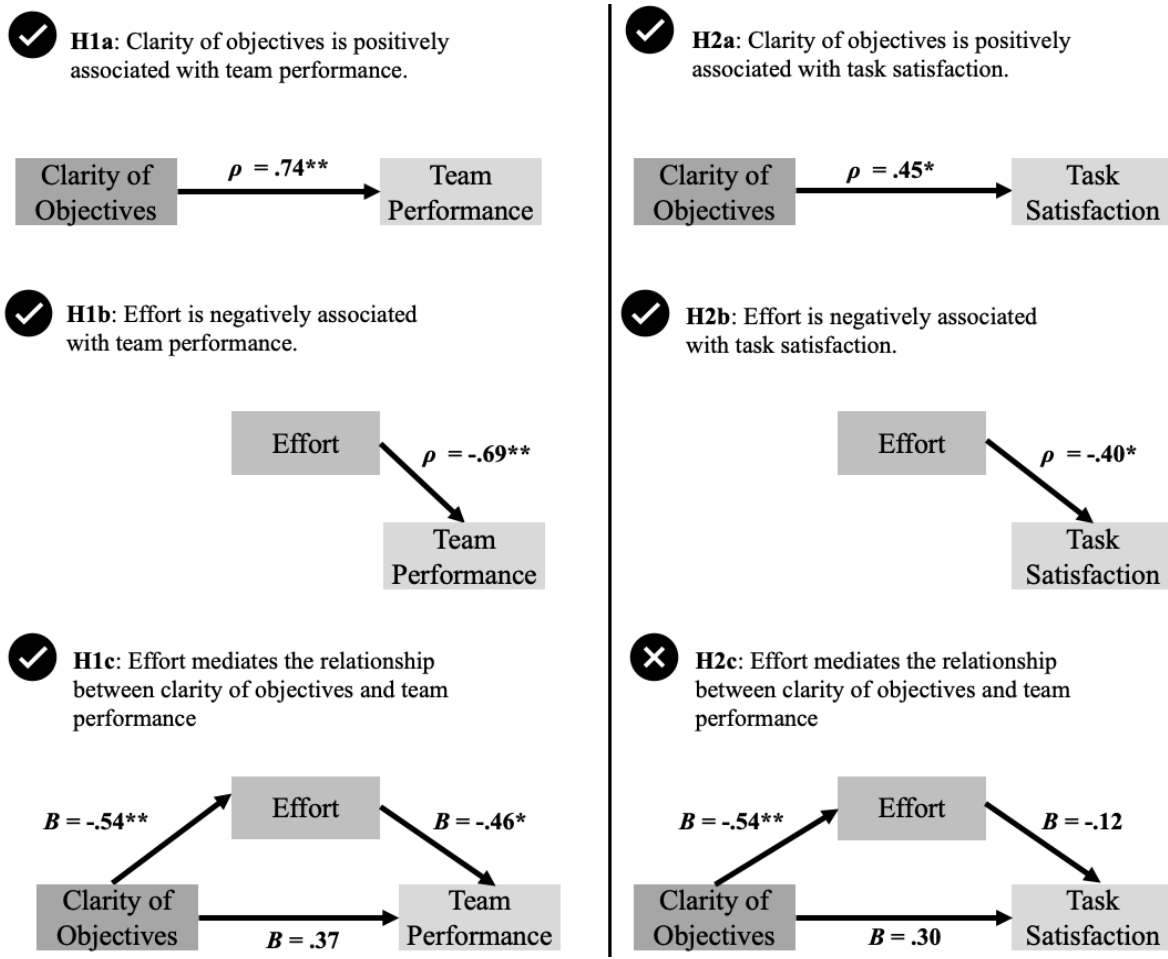

**Figure 4.**
*Overview of the results of the mediation analyses. Tick = Hypothesis supported; Cross = Hypothesis not supported;*
*\* = p <.05, \*\* = p <.01.*

H3a and H3b proposed that information flow is positively associated with team performance and frustration negatively. Table 1 shows a significant positive correlation between information flow and team performance ($\rho = .79$, $p < .01$), supporting H3a. Furthermore, the correlation between frustration and team performance is significant ($\rho = -.51$, $p < .05$), thus supporting H3b.

In H3c it was assumed that information flow reduces experienced frustration and this in return enhances team performance. The results of the mediation analysis showed that there was a total effect of information flow on team performance ($B = .68$, $p < .01$) in the assumed direction. After including the mediator into the model, information flow predicted frustration significantly and negatively ($B = -.72$, $p < .01$), but frustration in return did not predict team performance significantly ($B = .12$, $p > .05$, indirect effect $ab = -.09$, 95%-CI[-.88, .28]; see Appendix B, Tables B.9 to B.13 for a detailed overview of the results). Therefore, H3c was not supported (see Figure 5).

H4a and H4b proposed that information flow is positively associated with task satisfaction and frustration negatively. The correlation analyses in Table 1 show that there is a significant positive correlation between information flow and task satisfaction ($\rho = .58$, $p < .05$), supporting H4a. The correlation between frustration and task satisfaction is significantly negative ($\rho = -.57$, $p < .01$), supporting H4b.

It was expected in H4c that information flow reduces experienced frustration and this in turn enhances task satisfaction. The results of the mediation analysis showed that there was a total effect of information flow on task satisfaction ($B = .62$, $p < .05$) in the assumed direction. After including the mediator into the model, information flow significantly predicted frustration negatively ($B = -.72$, $p < .01$), and frustration in return predicted task satisfaction significantly and negatively ($B = -.69$, $p < .05$, indirect effect $ab = .49$, 95%-CI[.29, .91]; see Appendix B, Tables B.10 to B.11 and Tables B.14 to B.16 for a detailed overview of the results). The relationship between information flow and task satisfaction was not significant after the mediator was entered. Therefore, it was shown that the relationship between information flow and task satisfaction was mediated by frustration, supporting H4c (see Figure 5).

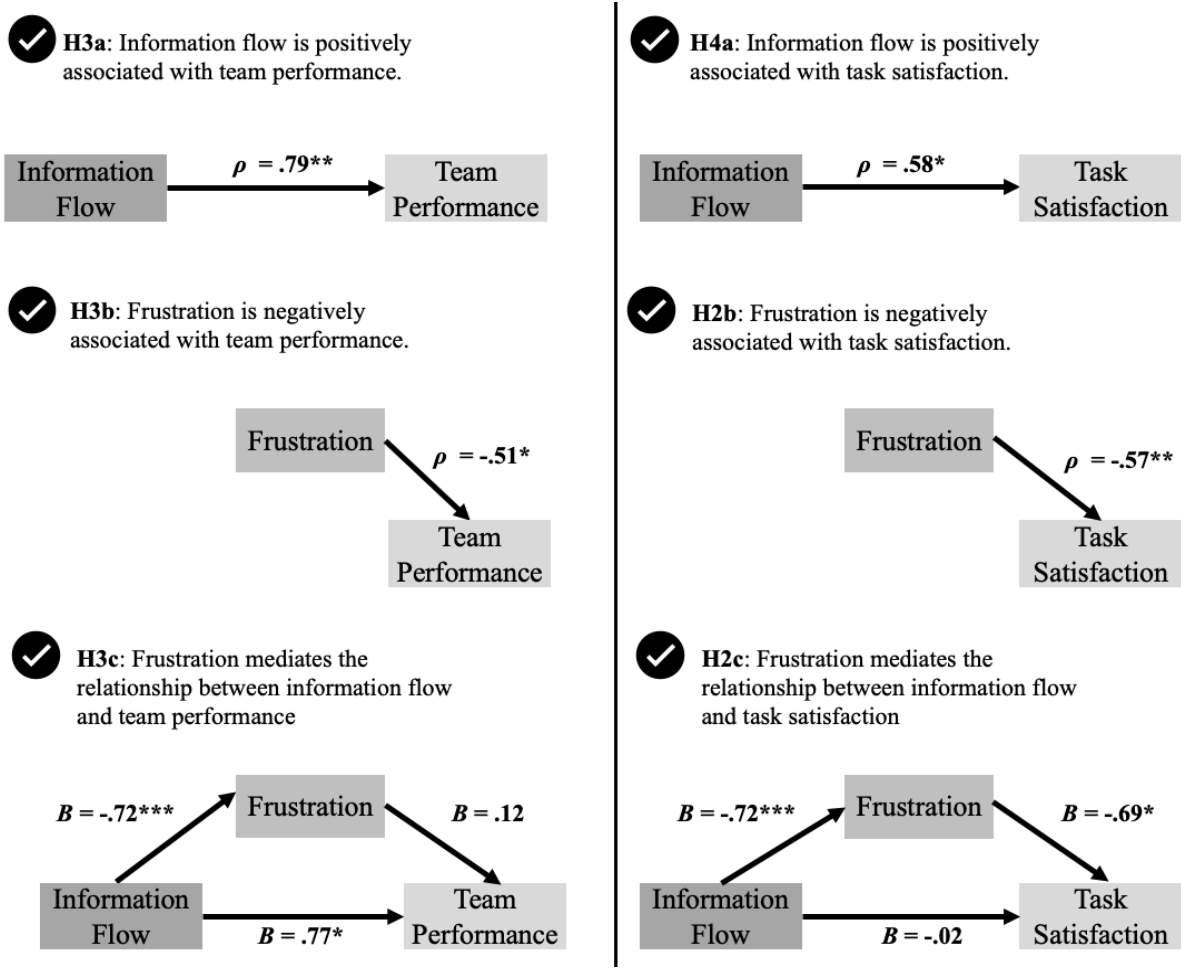

**Figure 5.**
*Overview of the results of the mediation analyses. Tick = Hypothesis supported; Cross = Hypothesis not supported;*
*\* = p <.05, \*\* = p <.01, \*\*\* = p <.001.*

## 4. Discussion and Conclusion

In future long-duration spaceflight missions, such as missions to Mars, the team of astronauts need to cope with the challenges of communication delays and even possible blackouts of communication. The context of spaceflight has shown that mistakes in the communication and malfunctioning teamwork had severe, even life-threatening consequences [36]. Therefore, it becomes increasingly important when preparing for future long-duration spaceflight missions to focus on the one hand on the teamwork within the team of astronauts, but on the other hand also on the teamwork of astronauts with their support teams considering the time delay in their communication [37].

This study adds to the existing research in the context of real and analog spaceflight by taking time-delayed communication and its impact on teamwork and also task satisfaction in a multiteam system into account. It was shown that there is a positive relationship between both facets of communication quality that have shown to be important in time-delayed communication, namely clarity of objectives and information flow, with objectively assessed team performance and subjectively assessed task satisfaction of the team members. These findings follow up on Kintz's et al. [21] research who showed that information flow is important to achieve successful team outcomes. Additionally, the study extends the findings of Fischer and Mosier [17] and Olson and Olson [25] who demonstrated that a common understanding, which is achieved through clarifying the objectives, is necessary to accomplish successful team performance. Furthermore, it was expected in the present study that the relationship between clarity of objectives and team performance is mediated by effort. This was also supported in the results indicating that when the goals are clear and a mutual understanding has been established, the mental effort is reduced, which in return results in a higher team performance. These results now combine single findings from previous research as it was shown that establishing a common understanding is more effortful in time-delayed communication due to missing immediate feedback and a high cognitive load [17,18,19]. The study further adds to previous research by showing that time-delayed communication not only leads to a high mental effort, but that this effort mediates the relationship between quality of communication, in terms of clarifying the objectives, and team performance. It was also assumed that effort mediates the relationship between clarity of objectives and task satisfaction, which was not found in the results. One explanation for this finding could be that task satisfaction was assessed from a subjective perspective (self-disclosure) and is also an individual attribute, which might not necessarily be affected by mental effort. Since astronauts and their support teams belong to HRTs they experience a high cognitive load in their tasks on a regular basis [38]. Therefore, it is possible that they are used to experiencing mental effort, so that it does not influence the experienced task satisfaction.

Against the expectation frustration did not mediate the relationship between information flow and team performance. As unsuccessful team performance of astronauts and their support teams leads to severe consequences [8,9], these HRTs place the highest value on being successful as a team. A possible explanation for finding no mediation effect of frustration on information flow and team performance could be that regardless of whether the team members feel frustrated or not, successful team performance and professional behaviour is prioritised and therefore, not affected by frustration of the team members. Nevertheless, it was shown in the results that frustration mediates the relationship between information flow and task satisfaction indicating that the experienced frustration due to bad information flow affects the subjective task satisfaction. These findings are in line with research that showed that time-delayed communication affects the team's efficiency and coordination, which in return leads to experiencing more frustration and negatively impacted individual outcomes [21]. Overall, these findings demonstrate that time-delayed communication has a considerable effect not only on performance outcomes but also on the subjective perception of teamwork.

However, these findings need to be considered in the light of the following limitations. This study is based on a small sample size of six participants per multiteam system and a total sample of 18 participants. It is therefore questionable to what extent the results can be transferred to other teams. Moreover, the analog Mars mission AMADEE-20, in which this study was conducted, only lasted four weeks. When considering real long-duration spaceflight missions, the duration of those missions will be much longer than four weeks. Thus, long-term effects beyond one month of communication on team performance and task satisfaction could not be observed in this study. Nonetheless, it can be noted as a strength that this sample is representative for the context of spaceflight and that it is a rare exception to get access to such a team in the field and do research in the wild [39,40].

For future research it is advisable to replicate our findings to find further evidence for the relationships demonstrated in this study and, more important, how these relationships might change with increasing time. In addition, it would be interesting to explore the role of other variables, such as the autonomy of the astronauts, in relation to communication and its outcomes to be able to develop effective training programs for astronauts and their support teams in preparation for long-duration spaceflight. Especially autonomy is important to consider when investigating long-duration spaceflight as the time-delay forces astronauts to act more autonomously compared to missions in Low Earth Orbit where astronauts work under strict control by their ground support [20,41]. Thus, it is important to further examine how high levels of autonomy influence the communication to ground support and the team performance as well as task satisfaction of teams of astronauts. There are studies that show that with increasing duration of the mission and increasing level of autonomy, astronauts tend to communicate less with their ground support [37,42]. Nevertheless, the communication between astronauts and their ground support remains essential, also over long-duration spaceflight missions, and it is important to develop communication trainings for astronauts and their support teams to avoid misunderstandings and to communicate quickly and efficiently despite possible communication delays.

Dempsey and Barshi [41] developed different training principles for preparing astronauts and their support teams for future long-duration spaceflight missions in a crew-centred and mission-oriented approach. With regard to the communication of those teams, they highlight different training principles. One training principle relates to the strategic use of knowledge. They argue to use strategies based on information that those teams are already familiar with and to use, for example, common procedures to elicit this familiarity [41]. Based on this, one strand of a possible communication training could be to develop unified commands, which is already the case in spaceflight missions, and to develop clear communication procedures. Furthermore, due to the increasing autonomy of the astronauts, it would be advisable in those trainings to also train for using existing knowledge to build retrieval cues that can be used to cope with new and challenging information since the communication delay to ground support will force astronauts to decide on their own when time is limited [41,43]. Another training principle provided by Dempsey and Barshi [41] is feedback. This becomes more difficult when thinking about astronauts training during the mission, e.g., to update learned skills, as feedback is only transferred time delayed. It seems that even though there are training approaches for astronauts and their support teams in preparation for long-duration spaceflight missions, these training approaches do only marginally or not at all focus on the concrete communication of those teams or in particular of those multiteam systems [36,41]. Nevertheless, since the time-delayed communication to ground support is the only possible line of contact for the astronauts, research should further examine the findings of this study to develop trainings for successful time-delayed communication and team performance. Additionally, current research interests develop in the direction to develop artificial intelligence systems that support crews during long-duration spaceflight missions [44]. There are approaches that focus on developing artificial intelligence systems that train the field crew and can directly give feedback on the given performance of the crew. Even though this is not the same as real-time contact to ground support, these are promising approaches to develop systems supporting crews during long-duration spaceflight missions.

Following on from that, Croitoru et al. [36] explored challenges to traditional team training in relation to team training in long-duration spaceflight missions and set up implications for further developing team training in this field. They highlight as one challenge that on the one hand during long-duration spaceflight mission, no instructor is there to train with the astronauts, so that methods, such as video or simulation training need to be developed more advanced to use it during those mission [36]. On the other hand, they also outline that no immediate feedback can be given to the astronauts. In their implications for the development of team trainings in long-duration spaceflight missions, they emphasize that regarding communication a shared language is essential [36]. This underlines again that clarity of objectives is a highly important feature of communication and furthermore, that developing clear communication procedures is essential for successfully training astronauts and their support teams. That is also supported by Fischer and Mosier [1] who found that using protocol approaches in time-delayed communication is promising for fostering effective and successful collaboration and communication between astronauts and their support teams. They also argue that implementing protocol elements could possibly reduce the cognitive load experienced by both astronauts and their support teams due to the time-delayed communication [1], which could be supported in consideration of the results of this study.

**Summary**

This study highlighted the importance of two facets of communication quality, namely clarity of objectives and information flow, when communication is time delayed in the context of interdependent teamwork during an analog Mars mission. Results revealed that clarity of objectives positively influences the team performance and the perceived task satisfaction of team members working together as a multiteam system despite a 10-minutes time delay. Furthermore, it was demonstrated that mental effort mediates the relationship between clarity of objectives and team performance, showing that when the objectives are clearly communicated, less mental effort is required, so that team performance improves. Information flow also has a positive effect on team performance and task satisfaction. The relationship between information flow and task satisfaction was mediated by frustration, which highlights that a high information flow leads to less frustration in the team, which in return increases task satisfaction.

Based on these findings, recommendations for future research are to test these findings in missions that last longer than four weeks and the influence of other variables, e.g., increasing autonomy of the crew, to develop trainings that improve communication quality, in specific the facets clarity of objectives and information flow, under time delay. To be able to further investigate this topic, it would be desirable to set up missions with longer durations, to vary with time delay as real signal travel time between Earth and Mars is up to 20 minutes one-way, and to consider specific communication trainings in advance to facilitate communication under time delay.

**Declaration of competing interest**

The authors declare that they have no known competing personal or financial interests that could have influenced the work reported in this paper.

**Acknowledgements**

The authors would like to thank the Austrian Space Forum, the Israeli Space Agency and the organisation D-MARS for hosting the AMADEE-20 analog Mars mission. We would especially like to thank our direct contacts, namely Karin Brünnemann, Dr. Seda Özdemir-Fritz, Simone Paternostro, Sophie Gruber and many more. Additionally, we like to express our gratitude to all participants of our study and those that supported our study during the mission. In particular, we would like to thank the six analog astronauts for their outstanding commitment and their efforts.

This research did not receive any specific grant from funding agencies in the public, commercial, or not-for-profit sectors.

# Appendix A

## *A.1 Team Task: Solving a murder case*

**Table A.1**

*Overview of the given team task: to solve a murder case.*

| Team member | Riddle to solve |
|---|---|
| AAs | Name of murdered person |
| OSS | Date of murder |
| MSC | Cause of death |
| AAs | Relationship between murderer and murdered person |
| OSS | Motive of murder |
| MSC | Name of murderer |
| AAs | Prison where murderer was arrested |
| OSS | Date of prison outbreak |
| MSC | Combine all information |

## A.2 Communication Measure adapted by Hirst & Mann (2004) and Wauben et al. (2010)

**Table A.2**

*Overview of Items used for Communication Scale from Communication Measure by Hirst & Mann (2004).*

| Items from Communication Measure by Hirst & Mann (2004) | Items used for INTERTEAM |
|---|---|
| *Clarity of Objectives* | |
| Team members have a clear understanding of project objectives. | All of us in the team had a clear understanding of the task's objectives. (*Clarity of Objectives*) |
| *Information Flow* | |
| Information circulates throughout the team. | Information regarding the task circulated through the team. (*Information Flow*) |
| It is often difficult to get answers to important questions about my work. | It was easy to get answers to important task-related questions within the team. (*Information Flow*) |

**Table A.3**

*Overview of Items used for Communication Scale from Communication Measure by Wauben et al. (2010).*

| Items from *Three Categories of communication* Measure by Wauben et al. (2010) | Items used for INTERTEAM |
|---|---|
| C1 - Exchanging information | Knowledge and information were circulating in order to establish a shared understanding of the task among us. (*Information Flow*) |
| C2 - Establishing a shared understanding | I perceived a shared understanding of the task among all team members. (*Clarity of Objectives*) |
| C3 - Coordinating team activities | Our team coordinated the task activities in time in a simultaneous and collaborative manner. (*Information Flow*) |

*A.3 Task satisfaction Measure adapted by Spector's (1985) Measurement of Human Service Staff Satisfaction*

**Table A.4**

*Overview of Items used for Task satisfaction Scale from Measurement of Human Service Staff Satisfaction by Spector (1985).*

| Items from Measurement of Human Service Staff Satisfaction by Spector (1985) | Items used for INTERTEAM |
|---|---|
| My job is enjoyable. | The work during the task was enjoyable. |
| Work assignments are not fully explained. | It was easy for me to conduct the task. |
| I have too much to do at work. | The workload of the task was attainable during the given time span. |
| I like doing the things I do at work. | I felt satisfied after finishing the task. |

# Appendix B

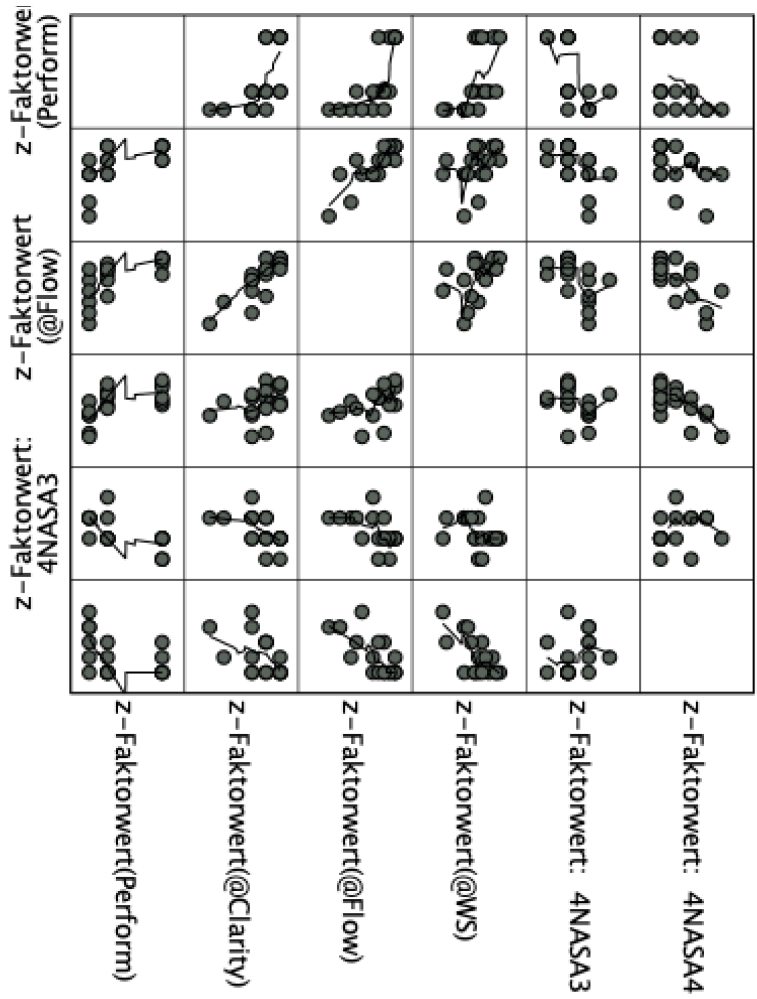

**Figure B. 1**
*Checking linearity by means of matrix diagrams and LOESS smoothing.*
*Note.* N = 18.

**Table B.1**
*Total effect of clarity of objectives on team performance (Model 1).*

| Variable | B | SE [a] | t | p | 95% CI |
|---|---|---|---|---|---|
| Clarity of objectives | .62 | .17 | 3.69 | .00** | [.26, .97] |

*Note.* $N = 18$. CI = confidence interval. Based on Bootstrapping (5000 iterations).

[a] Robust standard error.

** p < .01.

**Table B.2**
*Model summary of clarity of objectives on effort (Model 1 +2).*

| | R | $R^2$ | F (df1, df2) | p |
|---|---|---|---|---|
| Effort | .54 | .29 | 15.59 (1, 16) | .00** |

*Note.* $N = 18$. Based on Bootstrapping (5000 iterations).

** p < .01.

**Table B.3**
*Coefficients of clarity of objectives' regression analysis on effort (Model 1 + 2).*

| Variable | B | SE [a] | t | P | 95% CI |
|---|---|---|---|---|---|
| Constant | .00 | .20 | .00 | 1.00 | [-.43, .43] |
| Clarity of objectives | -.54 | .14 | -3.95 | .00** | [-.82, -.27] |

*Note.* $N = 18$. Based on Bootstrapping (5000 iterations).

** p < .01.

**Table B.4**
*Model summary of effort on team performance (Model 1).*

| | R | $R^2$ | F (df1, df2) | p |
|---|---|---|---|---|
| Team performance | .73 | .53 | 12.50 (2, 15) | .000*** |

*Note.* $N = 18$. Based on Bootstrapping (5000 iterations). CI = confidence interval.

*** p < .001.

**Table B.5**
*Coefficients of effort's regression analysis on team performance (Model 1).*

| Variable | B | SE [a] | t | p | 95% CI |
|---|---|---|---|---|---|
| Constant | .00 | .17 | .00 | 1.00 | [-.37, .37] |
| Clarity of objectives | .37 | .18 | 2.10 | .05 | [-.01, .75] |
| Effort | -.46 | .19 | -2.46 | .03* | [-.85, -.06] |

*Note.* N = 18. Based on Bootstrapping (5000 iterations). CI = confidence interval.

a Robust standard error.

* p < .05.

**Table B.6**
*Total effect of clarity of objectives on task satisfaction (Model 2).*

| Variable | B | SE [a] | t | p | 95% CI |
|---|---|---|---|---|---|
| Clarity of objectives | .37 | .15 | 2.42 | .03* | [.05, .69] |

*Note.* N = 18. CI = confidence interval. Based on Bootstrapping (5000 iterations).

[a] Robust standard error.

* p < .05.

**Table B.7**
*Model summary of effort on task satisfaction (Model 2).*

| | R | $R^2$ | F (df1, df2) | p |
|---|---|---|---|---|
| Task satisfaction | .38 | .15 | 3.12 (2, 15) | .07 |

*Note.* N = 18. Based on Bootstrapping (5000 iterations). CI = confidence interval.

**Table B.8**
*Coefficients of effort's regression analysis on task satisfaction (Model 2).*

| Variable | B | SE [a] | t | p | 95% CI |
|---|---|---|---|---|---|
| Constant | .00 | .23 | .00 | 1.00 | [-.49, .49] |
| Clarity of objectives | .30 | .24 | 1.27 | .22 | [-.21, .81] |
| Effort | -.12 | .28 | -.43 | .67 | [-.72, -.48] |

*Note.* N = 18. Based on Bootstrapping (5000 iterations). CI = confidence interval.

a Robust standard error.

**Table B.9**
*Total effect of information flow on team performance (Model 3).*

| Variable | B | SE [a] | t | p | 95% CI |
|---|---|---|---|---|---|
| Information flow | .68 | .15 | 4.64 | .00*** | [.37, .99] |

*Note.* $N = 18$. CI = confidence interval. Based on Bootstrapping (5000 iterations).

[a] Robust standard error.

*** $p < .001$.

**Table B.10**
*Model summary of information flow on frustration (Model 3+4).*

| | R | $R^2$ | F (df1, df2) | p |
|---|---|---|---|---|
| Frustration | .72 | .51 | 34.17 (1, 16) | .00*** |

*Note.* $N = 18$. Based on Bootstrapping (5000 iterations).

*** $p < .001$.

**Table B.11**
*Coefficients of information flow's regression analysis on frustration (Model 3+4).*

| Variable | B | SE [a] | t | P | 95% CI |
|---|---|---|---|---|---|
| Constant | .00 | .17 | .00 | 1.00 | [-.36, .36] |
| Information flow | -.72 | .12 | -5.85 | .00*** | [-.98, -.46] |

*Note.* $N = 18$. Based on Bootstrapping (5000 iterations).

*** $p < .001$.

**Table B.12**
*Model summary of frustration on team performance (Model 3).*

| | R | $R^2$ | F (df1, df2) | p |
|---|---|---|---|---|
| Team performance | .69 | .47 | 9.83 (2, 15) | .00** |

*Note.* $N = 18$. Based on Bootstrapping (5000 iterations). CI = confidence interval.

** $p < .01$.

**Table B.13**
*Coefficients of frustration's regression analysis on team performance (Model 1).*

| Variable | B | SE [a] | t | p | 95% CI |
|---|---|---|---|---|---|
| Constant | .00 | .18 | .00 | 1.00 | [-.39, .39] |
| Information Flow | .77 | .28 | 2.76 | .01* | [.18, 1.36] |
| Frustration | .12 | .31 | .38 | .71 | [-.55, .79] |

*Note. N* = 18. Based on Bootstrapping (5000 iterations). CI = confidence interval.
a Robust standard error.
* p < .05.

**Table B.14**
*Total effect of information flow on task satisfaction (Model 4).*

| Variable | B | SE [a] | t | p | 95% CI |
|---|---|---|---|---|---|
| Information flow | .48 | .16 | 3.08 | .01* | [.15, .81] |

*Note. N* = 18. CI = confidence interval. Based on Bootstrapping (5000 iterations).
[a] Robust standard error.
* p < .05.

**Table B.15**
*Model summary of frustration on task satisfaction (Model 4).*

| | R | $R^2$ | F (df1, df2) | p |
|---|---|---|---|---|
| Task satisfaction | .68 | .46 | 10.40 (2, 15) | .00** |

*Note*. N = 18. Based on Bootstrapping (5000 iterations). CI = confidence interval.
** p < .01.

**Table B.16**
*Coefficients of frustration's regression analysis on task satisfaction (Model 4).*

| Variable | B | SE [a] | t | p | 95% CI |
|---|---|---|---|---|---|
| Constant | .00 | .18 | .00 | 1.00 | [-.39, .39] |
| Information Flow | -.01 | .29 | -.06 | .95 | [-.63, .59] |
| Frustration | -.69 | .32 | -2.16 | .05* | [-1.37, -.01] |

*Note. N* = 18. Based on Bootstrapping (5000 iterations). CI = confidence interval.
a Robust standard error.
* p < .05.